# Reflective Metal/Semiconductor Tunnel Junctions for Hole Injection in AlGaN UV LEDs

Yuewei Zhang,[1,a)] Sriram Krishnamoorthy,[1] Fatih Akyol,[1] Jared M. Johnson,[2] Andrew A. Allerman,[3] Michael W. Moseley,[3] Andrew M. Armstrong,[3] Jinwoo Hwang,[2] and Siddharth Rajan[1,2,a)]

[1] Department of Electrical and Computer Engineering, The Ohio State University, Columbus, Ohio, 43210, USA

[2] Department of Materials Science and Engineering, The Ohio State University, Columbus, Ohio, 43210, USA

[3] Sandia National Laboratories, Albuquerque, New Mexico 87185, USA

**Abstract:** In this work, we investigate the use of nanoscale polarization engineering to achieve efficient hole injection from metals to ultra-wide band gap AlGaN, and we show that UV-reflective aluminum (Al) layers can be used for hole injection into p-AlGaN. The dependence of tunneling on the work function of the metal was investigated, and it was found that highly reflective Al metal layers can enable efficient hole injection into p-AlGaN, despite the relatively low work function of Al. Efficient tunneling hole injection was confirmed by light emission at 326 nm with on-wafer peak external quantum efficiency and wall-plug efficiency of 2.65% and 1.55%, respectively. A high power density of 83.7 W/cm$^2$ was measured at 1200 kA/cm$^2$. The metal/semiconductor tunnel junction structure demonstrated here could provide significant advantages for efficient and manufacturable device topologies for high power UV emitters.

---
a) Authors to whom correspondence should be addressed.
   Electronic mail: zhang.3789@osu.edu, rajan@ece.osu.edu

III-Nitride ultra-violet light emitting diodes (UV LEDs) are optimal sources for UV light due to several advantages including compact size, capability of ultra-fast operation and low environmental impact. Low cost, high efficiency solid state UV sources could replace the current technology (gas-based lamps), as well as enable new applications such as air disinfection and water sterilization.[1] However, widespread adoption of UV LEDs has been limited by the poor device efficiency.[2-9] Resistive p-type contact and poor p-type conduction remain as major challenges for achieving high efficiency UV LEDs.[10,11] A heavily doped, thick p-GaN layer is typically grown on top of the p-AlGaN cladding layer to enable Ohmic contact and hole injection. However, both the p-GaN layer and the p-type contact metal (typically a Ni- or Pt-based metal stack) can cause severe internal light absorption, contributing to reduced light extraction efficiency.[10,11] In addition, devices suffer from high resistance during operation due to poor hole injection and transport in the p-type layers.[12] As a result, the wall-plug efficiency of UV LEDs with emission wavelength below 365 nm remains lower than 6%.[12,13]

An alternative method for hole injection was proposed recently using a tunnel-injected UV LED structure with *UV-transparent n-AlGaN window layers* to avoid the challenges associated with direct p-type contact.[14,15] It enables n-type contacts for both bottom and top contact layers by connecting a transparent n-AlGaN top contact layer to the p-AlGaN cladding layer using an ultra-thin (< 4 nm) InGaN layer as shown in Fig. 1(a). The structure is expected to minimize internal light absorption and at the same time provide efficient hole injection. Using the tunnel-injected UV LED structure, efficient light emission at 325 nm was achieved with on-wafer peak external quantum efficiency of 3.37%.[15] Tunneling hole injection has also been achieved in higher composition p-$Al_{0.75}Ga_{0.25}N$, with 257 nm wavelength light emission demonstrated.[16]

Molecular beam epitaxy (MBE) provides optimal conditions for realizing tunnel-injected LEDs[17-20] with n-AlGaN contacts, since no p-AlGaN activation is required after growth, and relatively low growth temperatures for AlGaN prevents excessive decomposition or intermixing in InGaN layers. However, MBE is currently not a major production technology for commercial LEDs. In comparison, metalorganic chemical vapor deposition (MOCVD) serves as the most widely used technique for manufacturing UV LEDs, but there are significant challenges if InGaN-based tunnel junction LEDs with transparent window layers are to be realized. The n-AlGaN top contact layer impedes hydrogen diffusion, making it difficult to achieve Mg activation.[21] Even though lateral Mg activation from etched sidewalls has been demonstrated for MOCVD-grown tunnel-injected blue LEDs, they show increases in both the turn-on voltage and the differential resistance as compared to standard LEDs without tunnel junctions.[21,22] In addition, the high growth temperature needed for AlGaN makes it difficult to grow AlGaN layers (typically at ~ 1200 °C)[23] on top of the thin InGaN layer (typically grown at ~ 750 °C)[24] since the InGaN layer is subject to severe decomposition as well as intermixing when the growth temperature is increased for AlGaN growth. In this work, we demonstrate polarization engineered metal/InGaN/p-AlGaN tunnel junctions that could help realize MOCVD-based tunnel-injected UV LEDs. Based on the metal/semiconductor tunnel junction structure, we show that aluminum, as a highly UV reflective metal, can be directly used for p-type contact regardless of the low work function.

The concept demonstrated in this paper is to replace the n-AlGaN/InGaN/p-AlGaN tunnel junction structure with a metal/InGaN/p-AlGaN junction. The energy band diagrams for both structures are shown in Fig. 1. Because of the high density polarization sheet charge at the InGaN/AlGaN interface, strong polarization field builds up in the InGaN layer for both tunnel junction structures.[25-27] This leads to band alignment between the metal and the p-AlGaN valence band within short distance (< 4 nm). The interband tunnel barrier height is determined by InGaN bandgap for the n-AlGaN/InGaN/p-AlGaN semiconductor tunnel junction.[25,28] In comparison, the metal/InGaN/p-AlGaN tunnel junction can provide flexibility in the tunnel barrier height by varying metal work function.

Ohmic contacts using p-GaN capping on top of p-AlGaN or p-InGaN capping on top of p-GaN have been well-studied.[10,29,30] However, those p-type contacts are demonstrated inevitably with high work function metal layers, such as Ni- or Pd-based metal stacks, which have low reflectivity for UV light. The feasibility of using UV-reflective Al-based metal stack for p-type contact has not been studied. Here, the equilibrium energy band diagrams of the metal/semiconductor tunnel junction structure with Ni- and Al-based contacts are compared in Fig. 1(b). We assume there is no surface pinning effect, and the band alignment was calculated by taking Ni and Al work functions ($W_m$) as 5.01 eV and 4.08 eV, respectively. Regardless of the large work function difference between the metal layers and the p-AlGaN layer ($W_{p\text{-}AlGaN}$ > 7.5eV), both Ni and Al contacts cause very minimal depletion in the p-AlGaN layer. This is because of the large polarization-induced band bending in the InGaN layer, which acts as an ultra-thin tunneling barrier for hole injection.[14,25] When the tunnel junctions are under reverse bias, electrons tunnel from the valence band of p-AlGaN across the thin InGaN barrier into the empty states above the metal Fermi level, and therefore, holes are injected into the p-AlGaN layer.

The tunnel barrier height ($\Phi_B$) is determined by the energy difference between metal Fermi level and InGaN valence band edge as $\Phi_B = \chi_{InGaN} + E_g - W_m$, where $\chi_{InGaN}$ and $E_g$ represent electron affinity and bandgap of the InGaN layer, respectively. Since Al has lower work function than does Ni, a higher tunnel barrier is expected for the Al/InGaN/p-AlGaN tunnel junction as shown in Fig. 1(b). As a result, higher reverse bias across the tunnel junction layer is required to achieve sufficient hole injection when Al contact is used as compared to the case with Ni contact. However, since Al is highly reflective (reflectivity > 90%)[31] to UV light over the whole wavelength range from 400 nm down to 200 nm, the Al/InGaN/p-AlGaN tunneling contact could significantly benefit the light extraction efficiency in UV LEDs. Such a structure also provides advantages for optical design for laser diodes. Since the optical modes have a node at the metal/semiconductor interface, the field intensity of the UV light is low near the surface of the semiconductor, thereby reducing the effects of absorption in the narrow bandgap InGaN tunnel junction. Placing the absorbing tunnel junction near a node of the optical mode is therefore a significant advantage of such a device over n-AlGaN/InGaN/p-AlGaN tunnel junctions where absorption in the thin InGaN layer can impact the optical losses.

The samples investigated in this work were grown by N$_2$ plasma assisted MBE on metal polar Al$_{0.3}$Ga$_{0.7}$N templates with threading dislocation density of $2 \times 10^9$ cm$^{-2}$. The epitaxial structure is optimized based on our previous studies and is shown in Fig. 2 (a).[14,15,32] It consists of n+ $Al_{0.3}Ga_{0.7}N$ bottom contact layer, three periods of 2.5 nm $Al_{0.2}Ga_{0.8}N$/ 7.5 nm $Al_{0.3}Ga_{0.7}N$ quantum wells (QWs)/ barriers, 1.5 nm AlN electron blocking layer (EBL), 50 nm graded p-AlGaN with Al content grading down from 75% to 30%, and 4 nm unintentionally doped $In_{0.25}Ga_{0.75}N$ capping layer. Immediately after the InGaN layer growth, the substrate temperature was ramped down to room temperature to avoid InGaN decomposition.

The LED devices were fabricated by mesa etching using inductively coupled plasma reactive ion etching (ICP-RIE), Ti/Al/Ni/Au bottom metal deposition and subsequent annealing at 750 °C, and deposition of top metal stack.[15] The influence of two different top metal stacks were studied by depositing Al(30 nm)/ Ni(30 nm)/ Au(150 nm)/ Ni(20 nm) contact to one region of the UV LED sample, and Ni(30 nm)/ Au(150 nm)/ Ni(20 nm) contact to another region of the same sample to avoid sample to sample variation. To further investigate the effect of InGaN layer thickness on device electrical performance, low power plasma etch with an etch rate of 4.5 nm/min was used to recess the InGaN layer before top metal contact deposition. Based on the etch time, the remaining InGaN layer thicknesses were estimated to be 3.2, 2.0, 0.9, 0.2 nm in different devices. A same Al/Ni/Au/Ni top metal stack was then evaporated on those devices to form top contact.

Fig. 2(b) shows the high-angle annular dark-field scanning transmission electron microscopy (HAADF-STEM) image of the tunnel-injected UV LED device with Al-based contact. Smooth interfaces are observed for QWs, EBL and the InGaN tunneling layer. This indicates that the device fabrication process did not cause noticeable material degradation of the InGaN layer. The contrast gradient in the p-AlGaN layer reflects effective Al-compositional grading, which is critical to the formation of the three-dimensional negative polarization charge.[15,33] The graded p-AlGaN layer leads to a flat valence band for hole transport, but contributes to a high barrier to block electron overflow as shown in the equilibrium energy band diagram with Al top contact in Fig. 2(c). This is beneficial for enhanced carrier injection efficiency. Even though Al has a work function close to InGaN electron affinity, the sharp band bending in the InGaN layer aligns Al Fermi level to the p-AlGaN valence band edge within a short distance (< 4 nm). Thus, holes can be tunnel-injected into the p-AlGaN layer by applying positive bias on the Al top contact layer.

The current-voltage (IV) characteristics of 10×10 μm$^2$ micro-LEDs with different top contact metal stacks are shown as solid lines in Fig. 3(a). The voltages at 20 A/cm$^2$ are 4.63 V and 5.85 V for the devices with Ni- and Al-based contact, respectively. The higher operation voltage for the UV LEDs with Al top contact is expected to be a result of increased tunnel barrier height as shown in Fig. 1(b). The voltage difference increases from 1.22 V at 20 A/cm$^2$ to 2.05 V at 1 kA/cm$^2$. The differential resistances are $6.9 \times 10^{-4}$ Ω cm$^2$ and $7.7 \times 10^{-4}$ Ω cm$^2$ at 1 kA/cm$^2$ for the devices with Ni- and Al-based contacts, respectively. For comparison, the electrical characteristics of an identical UV LED with an n-AlGaN/InGaN/p-AlGaN semiconductor tunnel junction layer are shown in Fig. 3 as dashed lines. It shows higher turn-on voltage and differential resistances as compared to the devices with Al/InGaN/p-AlGaN tunnel junction even though similar InGaN tunnel barriers are expected as shown in Fig. 1. This is attributed to enhanced interband tunneling due to surface states at metal/semiconductor interface in the Al/InGaN/p-AlGaN tunnel junction structure.[7,34]

The electrical characteristics of the devices with different InGaN layer thicknesses are compared in Fig. 4. As the InGaN layer thickness is reduced, both the turn-on voltage and on-resistance increase. An abrupt increase from 6.35 V to 9.08 V occurs when the InGaN layer is reduced from 2.0 nm to 0.9 nm. This indicates a sharp increase in the tunnel barrier for hole injection, which might correspond to evident depletion in the p-AlGaN layer. The dramatic increase in the forward voltage from 5.85 V (4 nm InGaN) to 9.91 V (0.2 nm InGaN) at 20 A/cm$^2$ clearly demonstrates that the InGaN layer enhances tunneling significantly by reducing the tunnel barrier so that Al-based contact can be used for hole injection.

The electroluminescence (EL) spectrum of the device with Al-based top contact is shown in Fig. 5(a). It shows single peak emission with a blue shift of the peak wavelength from 328.3 nm to 325.1 nm with increasing injection current due to the quantum confined Stark effect.[35] The microscope image indicates efficient light emission from the tunnel-injected UV LED device. The emission power of the devices was measured on-wafer under continuous-wave operation. The devices with Al-based top contact exhibited higher power and efficiency as shown in Fig. 5 and Fig. 6. At the injection current of 1200 A/cm$^2$, high emission power density of 83.7 W/cm$^2$ and 49.2 W/cm$^2$ were measured for the devices with Al- and Ni-based top contact, respectively. The peak external quantum efficiency and wall-plug efficiency are 2.65% and 1.55% for the device with Al top contact, and 1.42% and 1.00% for the device with Ni top contact. The device with Al top contact exhibited 87% and 55% increases in the peak external quantum efficiency and wall-plug efficiency respectively as compared to the device with Ni top contact. This is attributed to higher light extraction efficiency associated with the high UV reflectivity of the Al top contact layer. However, a sharper efficiency droop was observed for the devices with Al top contact. This might be a result of increased heating effect[36] because of the higher voltage drop across the device as shown in Fig. 3.

One challenge associated with these devices was the failure of larger area devices at current densities exceeding ~ 20 A/cm$^2$. It was found that for device areas in excess of 30×30 µm$^2$, devices had catastrophic failure with increased leakage and no light emission once the current density was increased above a critical level. Since this was only observed in larger area devices, we attribute it to leakage through specific dislocation types, or to macroscopic defects in the sample. While further investigation is needed to determine the exact origin of this failure, the performance of smaller area devices serves as a proof-of-concept for the idea of polarization-enhanced metal/semiconductor tunnel junctions for UV LEDs. This structure provides growth flexibilities for various growth methods, including MBE and MOCVD techniques. At the same time, it replaces the widely used absorbing p-type contact layers using Al-based tunneling contact, and overcomes issues related to absorption in laser diodes by placing the lower bandgap InGaN next to the reflective metal. Since Al is unique in having a high reflectivity above 90% for UV light, this structure could potentially lead to significant increase in the light extraction efficiency for the UV emitters, and be especially useful in laser diode applications.

In summary, we have demonstrated a tunnel-injected UV LED structure using a metal/InGaN/p-AlGaN tunnel junction for hole injection. We compared the influence of Ni- and Al-based top contact metal stacks on the device performance. Higher turn-on voltage and differential resistance was observed using Al-based contacts. This is attributed to higher tunnel barrier originating from the lower work function of Al. Nonetheless, the device with Al top contact exhibited 87% and 55% increases in the peak external quantum efficiency and wall-plug efficiency, respectively, as compared to the device with Ni top contact. Through tunneling hole injection, we achieved light emission at 326 nm with on-wafer peak EQE and WPE of 2.65% and 1.55%, respectively. A high power density of 83.7 W/cm$^2$ was measured at 1200 kA/cm$^2$. This work demonstrates the potential application of metal/semiconductor tunneling contact for hole injection towards high efficiency UV emitters.

Acknowledgement:

We acknowledge funding from the National Science Foundation (ECCS-1408416 and PFI AIR-TT 1640700), and the OSU TCO Accelerator Award. Sandia National Laboratories is a multi-mission laboratory managed and operated by Sandia Corporation, a wholly owned subsidiary of Lockheed Martin Corporation, for the U.S. Department of Energy's National Nuclear Security Administration under contract DE-AC04-94AL85000.

Figure captions:

Fig. 1 Schematic structures, charge distributions and equilibrium energy band diagrams of (a) n-AlGaN/InGaN/p-AlGaN tunnel junction (TJ), and (b) metal/InGaN/p-AlGaN tunnel junction. The effects of Ni- and Al-based metal layers are compared in the energy band diagram in (b). Higher tunnel barrier height is resulted when Al contact is used because of its lower work function.

Fig. 2 (a) Epitaxial stack, (b) HAADF-STEM image, and (c) equilibrium energy band diagram of the UV LED with metal/ semiconductor tunnel junction contact.

Fig. 3 (a) Current-voltage characteristics and (b) differential resistances of the 10×10 µm$^2$ tunnel-injected UV LED devices with different top metal contacts. The electrical characteristics of an identical UV LED with an n-AlGaN/InGaN/p-AlGaN semiconductor tunnel junction layer are shown as dashed lines for comparison.

Fig. 4 (a) I-V characteristics of the UV LEDs with Al/InGaN/p-AlGaN tunnel contact, where the InGaN layer thickness is varied by low power plasma etch. The change of voltage drop at 20 A/cm$^2$ with the InGaN layer thickness is shown in (b). A dramatic increase in the turn-on voltage is observed with reducing InGaN layer thickness below 2 nm.

Fig. 5 (a) Electroluminescence spectra, and (b) output power of 10×10 µm$^2$ tunnel-injected UV LED devices with different top metal contacts obtained on-wafer under continuous-wave operation. It shows single peak emission at ~ 326 nm. The inset to (a) is a microscope image of the device with Al-based top contact operated at 500 A/cm$^2$.

Fig. 6 (a) EQE, and (b) WPE of the 10×10 µm$^2$ tunnel-injected UV LED devices with different top metal contacts. The results were measured on-wafer under continuous-wave operation. The higher emission efficiency from the devices with Al-based top contact is attributed to enhanced light extraction due to high reflectivity of Al to the UV light.

(a) Semiconductor TJ  (b) Metal/semiconductor TJ

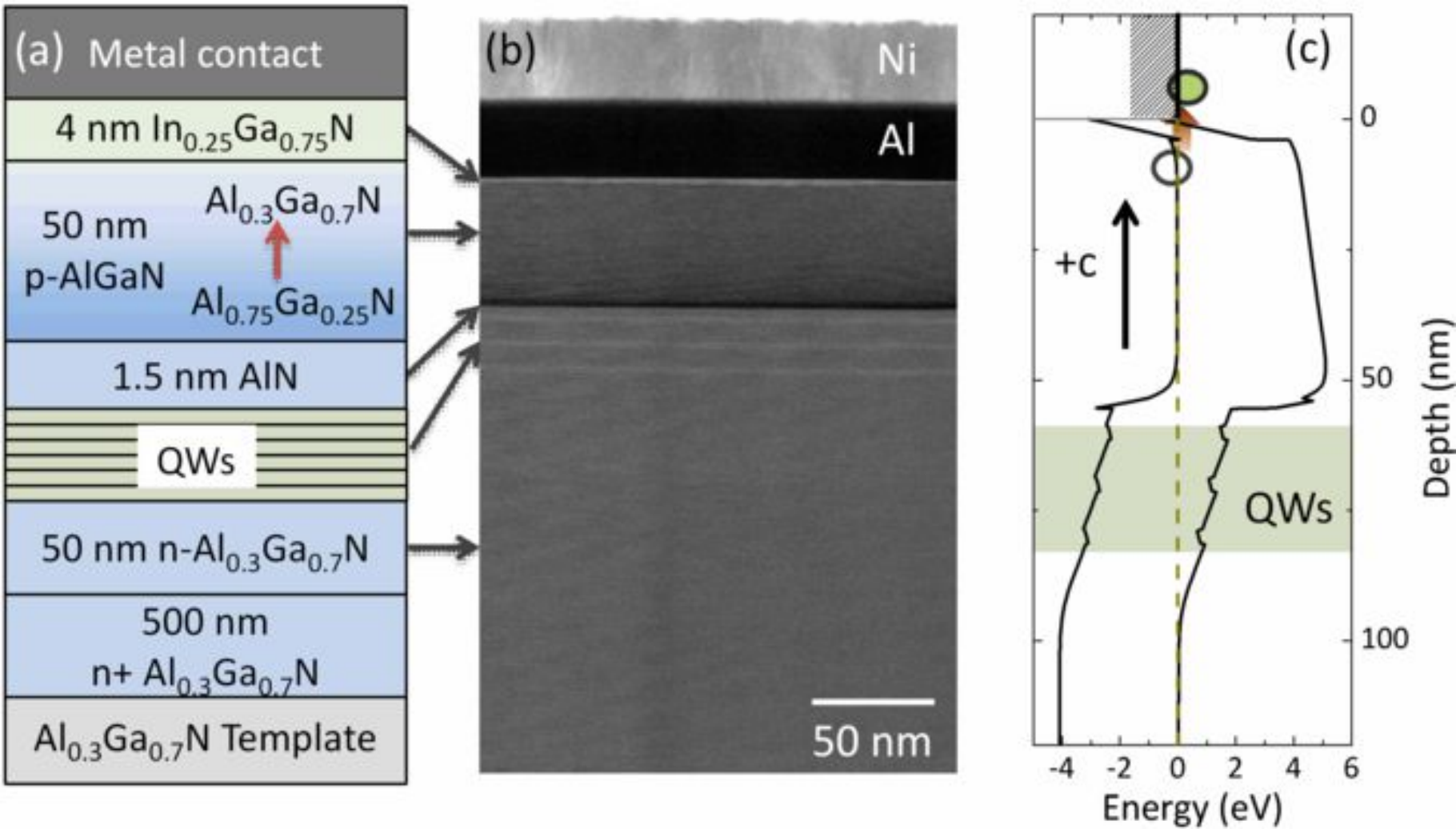

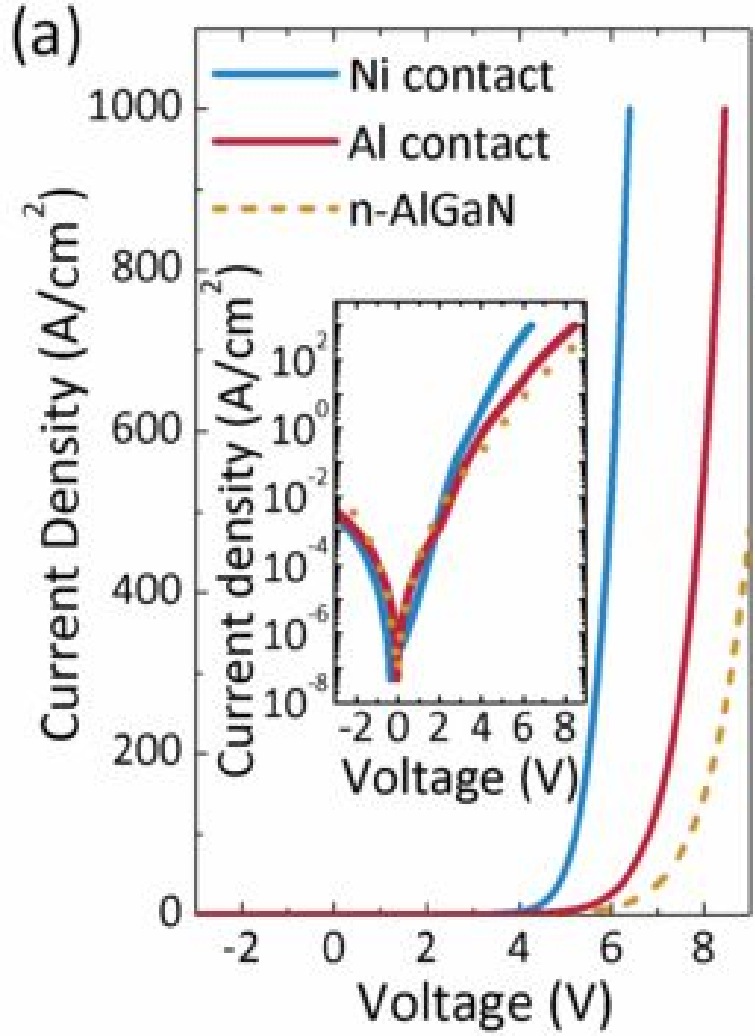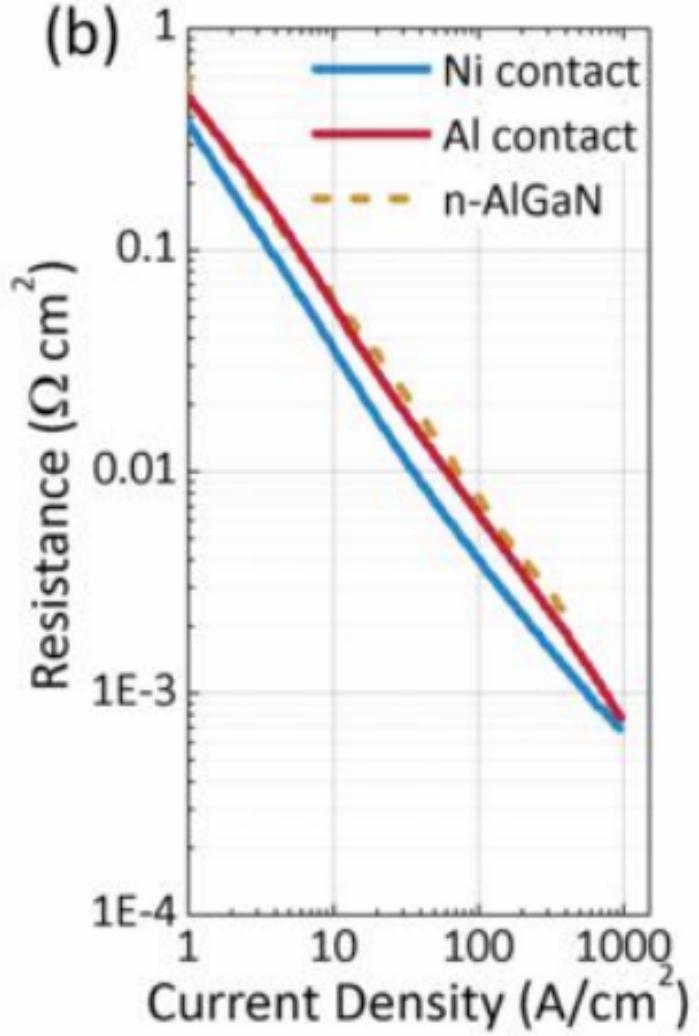

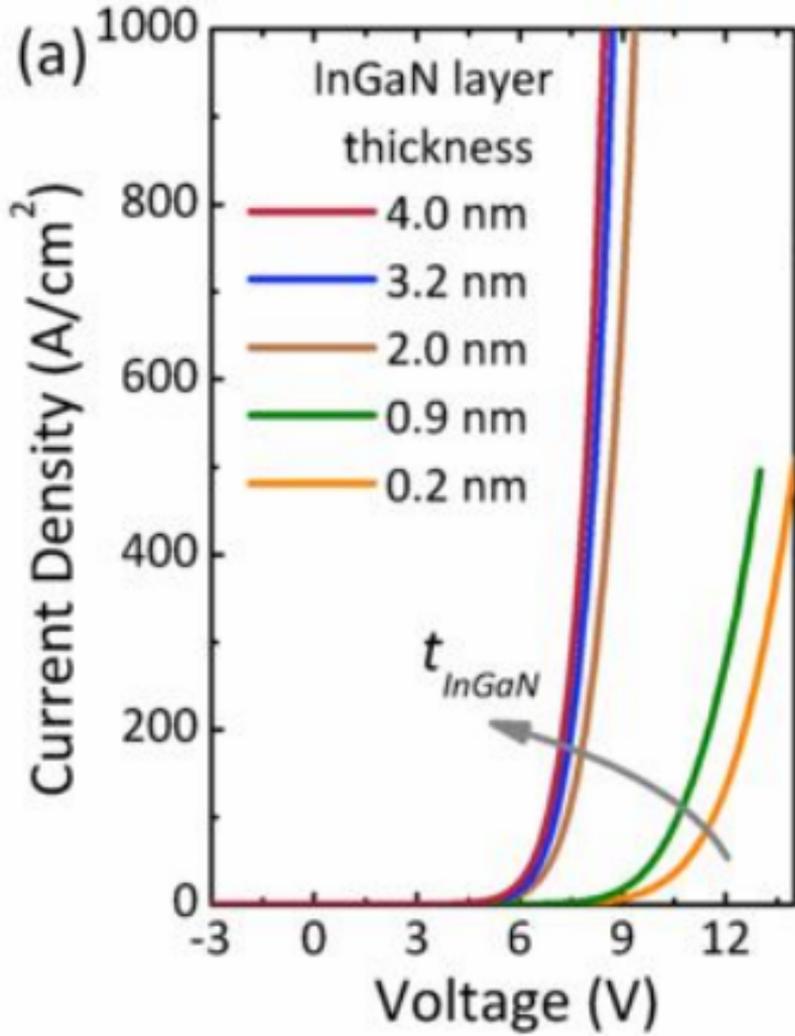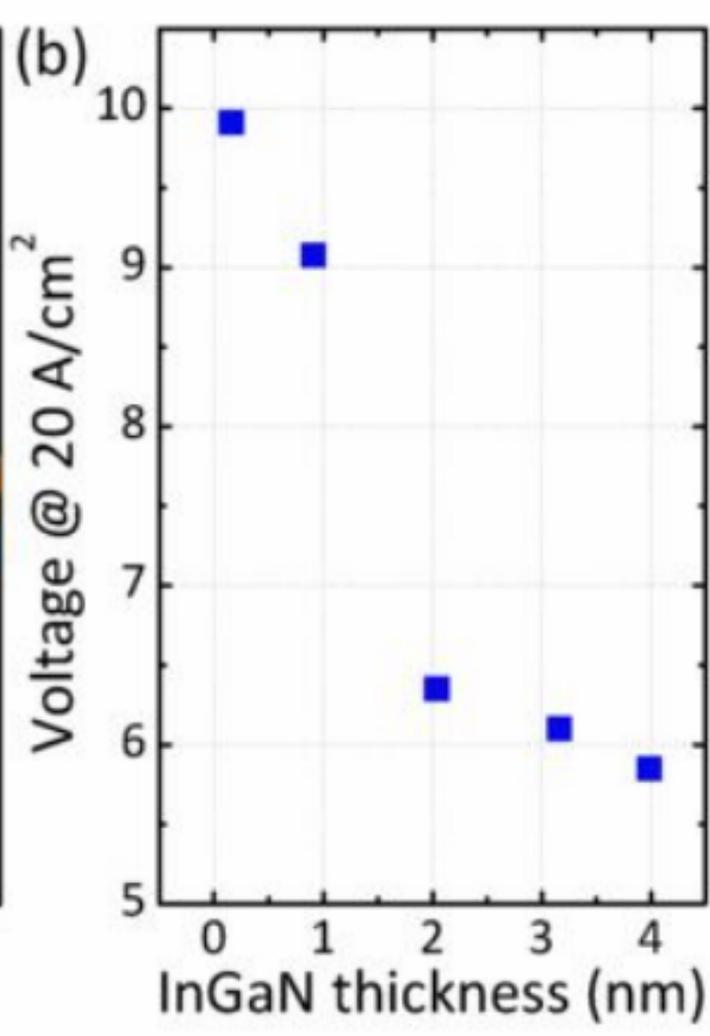

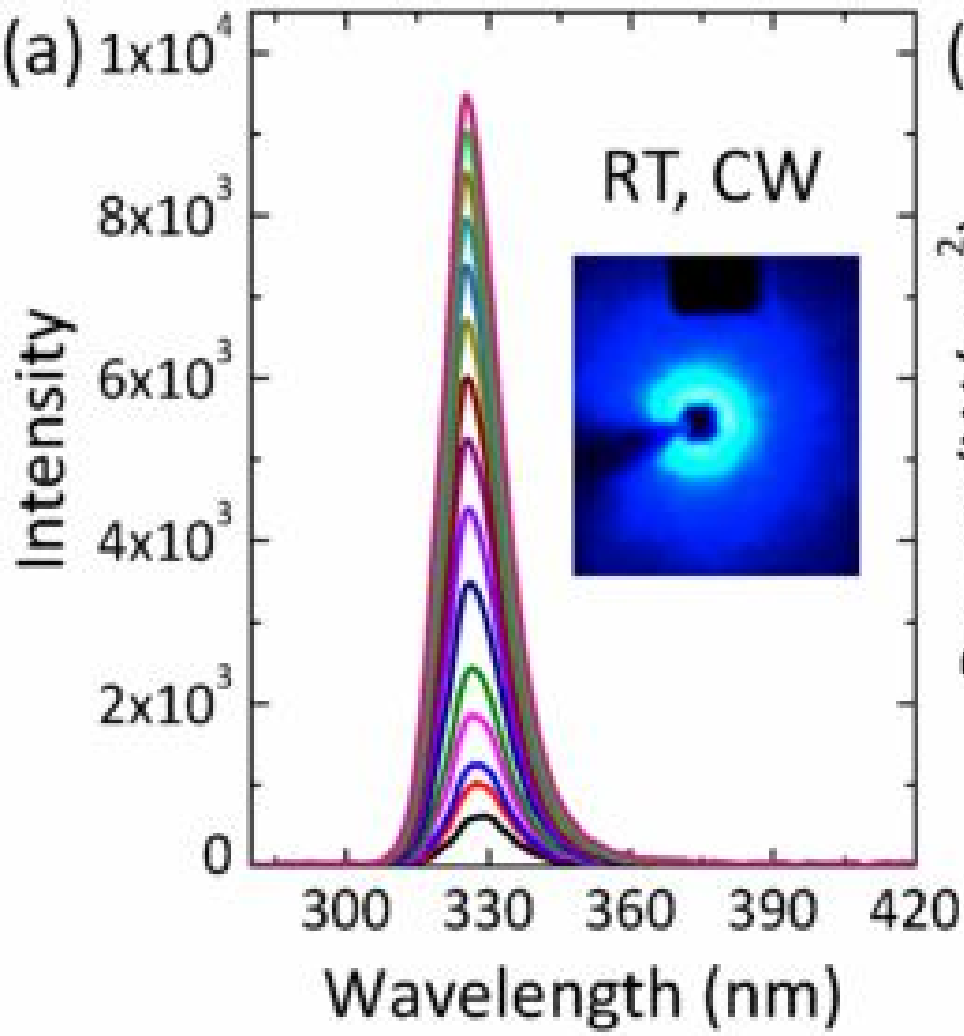 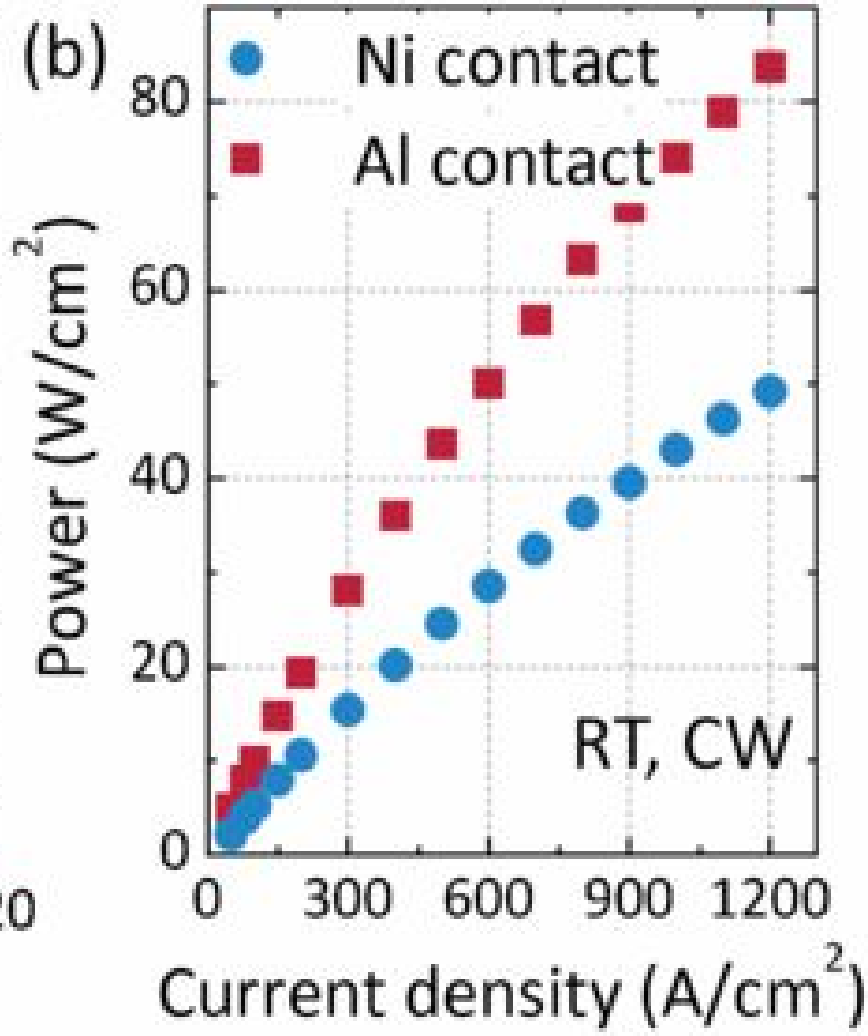

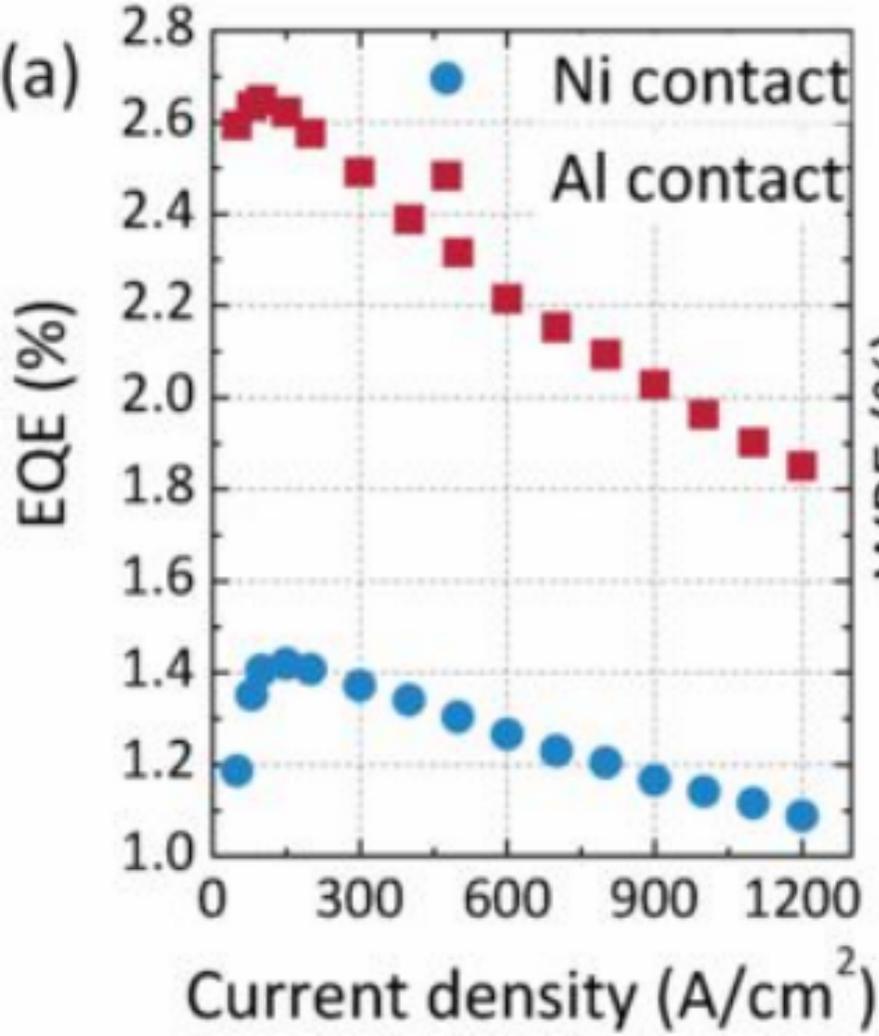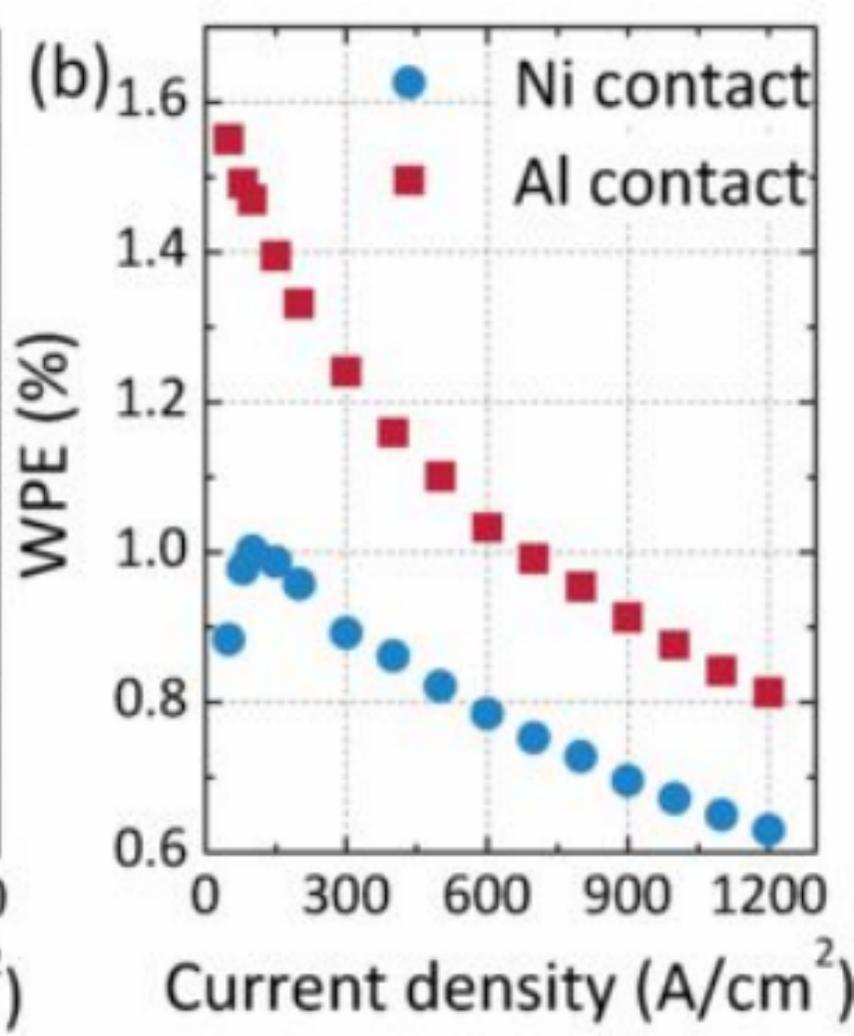